\documentclass[letterpaper, 10 pt, conference]{ieeeconf}  

\IEEEoverridecommandlockouts                              
\usepackage{amsmath} 
\usepackage{amssymb}  
\usepackage{mathptmx} 
\usepackage{subfigure}
\usepackage{mathrsfs} 
\usepackage{float} 
\usepackage{multirow} 
\usepackage{graphicx}
\usepackage{microtype} 
\usepackage{algorithm}
\usepackage{algpseudocode}

\algrenewcommand\alglinenumber[1]{#1:}
\usepackage{color}

\newtheorem{remark}{Remark}

\newcounter{rmnum}
\newenvironment{romannum}{\begin{list}{{\upshape (\roman{rmnum})}}{\usecounter{rmnum}
			\setlength{\leftmargin}{14pt}
			\setlength{\rightmargin}{8pt}
			\setlength{\itemsep}{2pt}
			\setlength{\itemindent}{-1pt}
	}}{\end{list}}

\newcounter{anum}

\def\qed{\ifmmode\IEEEQEDclosed\else{\unskip\nobreak\hfil
\penalty50\hskip1em\null\nobreak\hfil\IEEEQEDclosed
\parfillskip=0pt\finalhyphendemerits=0\endgraf}\fi}
\newcommand{\ud}{\,\mathrm{d}}

\def\v{{\sf K}}

\def\Re{\mathbb{R}}

\def\Expect{{\sf E}}

\def\clZ{{\cal Z}}

\newcommand{\dt}[1]{\frac{\ud{1}}{\ud t}}
\newcommand{\dthet}{\dot{q}}
\newcommand{\cm}{_{\text{\tiny CM}}}

\newcommand{\TT}{_{t}}
\newcommand{\NN}{_{n}}
\newcommand{\CC}{\mathsf{R}}
\newcommand{\fgen}{f^{\text{gen}}}
\def\v{{\sf K}}

\newcommand{\C}{C_{(q)}}
\renewcommand{\Re}{\mathbb{R}}
\newcommand{\xlc}{X^{LC}}

\newcommand{\HminN}{\underbar{H}^{(N)}}
\newcommand{\JN}{J^{(N)}}
\newcommand{\HN}{H^{(N)}}
\newcommand{\thetan}{\theta}
\newcommand{\thetaN}{\theta^{(N)}}
\newcommand{\cN}{c^{(N)}}
\newcommand{\hatHN}{\hat{H}^{(N)}}
\newcommand{\hatHminN}{\hat{\underbar{H}}^{(N)}}

\title{\LARGE \bf
	Bio-inspired Learning of  Sensorimotor Control for Locomotion
}

\author{Tixian Wang, Amirhossein Taghvaei, and Prashant G. Mehta
\thanks{Financial support from the ONR MURI grant N00014-19-1-2373 and the ARO grant W911NF1810334 is gratefully acknowledged.}
\thanks{T.~Wang, A.~Taghvaei and P.~G.~Mehta are with the 
    Coordinated Science Laboratory and the Department of 
    Mechanical Science and Engineering at the University of 
    Illinois at Urbana-Champaign (UIUC).
	{\tt\small tixianw2@illinois.edu; taghvae2@illinois.edu; mehtapg@illinois.edu}}
}

\begin{document}
	
\maketitle
\thispagestyle{empty}
\pagestyle{empty}

\begin{abstract}

This paper presents a bio-inspired central pattern generator
(CPG)-type 
architecture for {\em learning} optimal maneuvering control of periodic locomotory
gaits. The architecture is presented here with the aid of a snake robot
model problem involving planar locomotion of coupled rigid body
systems.  The maneuver involves clockwise or counterclockwise turning
from a nominally straight path.  The CPG circuit is realized as a
coupled oscillator feedback particle filter. The collective dynamics
of the filter are used to approximate a posterior distribution that is
used to construct the optimal control input for maneuvering the
robot.  A Q-learning algorithm is applied to learn the approximate
optimal control law.  The issues surrounding the parametrization of
the Q-function are discussed.  The theoretical results are illustrated
with numerics for a 5-link snake robot system.  


\end{abstract}

\section{INTRODUCTION}

\begin{figure}[t]
	\centering
	\includegraphics[width=0.95\columnwidth]{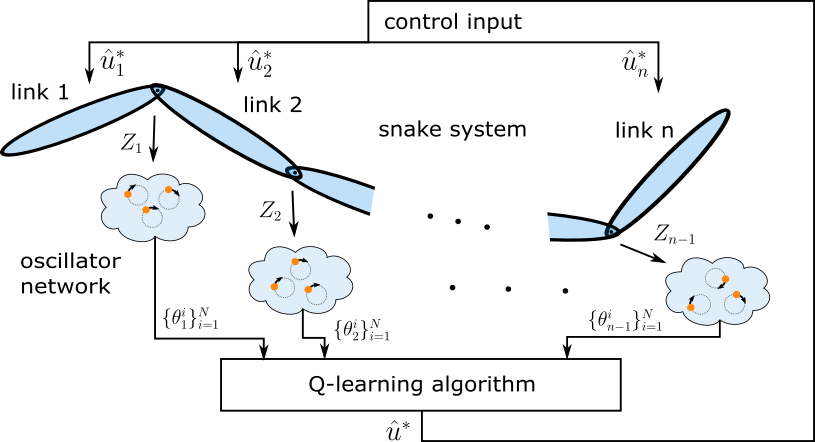}
	\caption{The proposed architecture to learn a distributed feedback control law for turning the snake robot.}
	\label{fig:ctrl_arch}
\end{figure}

The objective of this paper is to present a bio-inspired central
pattern generator (CPG)-type sensori-motor control architecture to
{\em learn} optimal maneuvers using {\em only} noisy sensor
measurements and (online) reward.  The dynamic and sensor
models are assumed unknown.    
The architecture, depicted in Fig.~\ref{fig:ctrl_arch}, is presented here
with the aid of a snake robot model problem involving planar
locomotion of coupled rigid body systems.  

The snake robot is modeled as
$n$ coupled rigid bodies.  The configuration 
space of the system is split into two sets of variables: 
(i) the {\it shape} variable which describes the internal shape of the system; 
(ii) and the {\it group} variable which describes the global displacement and 
orientation of the system.  
The shape variables are actuated using
motors at each joint to produce a nominal sinusoidal gait for the
forward motion.  The synthesis procedure for this gait is taken
from~\cite{IwasakiTAC2011_01}, where it was shown to be optimal with
respect to an energy cost function.  

The learning problem is for the robot to learn to maneuver
about this nominal gait.  The particular maneuver is to turn the robot
either clockwise or counter-clockwise, e.g., to avoid an obstacle in
the environment.  We assume noisy measurement of the shape variables
and employ changes in friction coefficients, with respect to the
surface, as control inputs.  

The main complexity reduction technique is to model the nominal
periodic motion of the (local) shape variable at the $j$-th joint
in terms of a single (hidden) phase variable $\theta_j(t)$ for
$j=1,2,\ldots,n-1$.    The inspiration comes from
	neuroscience where phase reduction is a popular technique to obtain
	reduced order model of neuronal
	dynamics~\cite{izhikevich2007dynamical}.

A coupled oscillator feedback particle filter (FPF) is used to approximate
the posterior distribution of $\theta_j(t)$ given noisy measurements.
The collective dynamics of the $(n-1)$ oscillator populations
electrically encode
the evolution of the mechanical shape of the robot.  
The filter requires knowledge of the observation model which is also
learned in an online fashion through the use of a linear
parameterization.  

The filter outputs are aggregated into the second layer which seeks to
learn the Q-function (or the Hamiltonian) based on an online access to
the reward.  A clever linear parametrization is used to enforce a
distributed architecture for the policy.  The parameters are
learned by using a 
	gradient descent algorithm to reduce the Bellman
        error~\cite{vrabie2007continuous,mehta2009q}.   

This overall control system can be viewed as a central pattern
generator (CPG) which integrates sensory information to learn
closed-loop optimal control  policies for bio-locomotion.  The
framework presented here is based upon our prior research in~\cite{taghvaei2014coupled} 
where phase reduction technique was introduced for a 2-link system and
in~\cite{tixian2019control} where the technique was extended to include learning for the
2-link system.  The main contributions of this work over and above
these prior publications are as follows:
\begin{enumerate}
\item The application involving the snake robot is new and more
  practically motivated than the simple 2-link model considered
  in~\cite{tixian2019control}.
\item The distributed coupled oscillator FPF is biologically
  motivated.  Each of the FPF encodes only the local shape and can be
  extended to $n$-links and ultimately to a continuum rod type
  models.  In contrast, the framework in our earlier papers
  parametrized the limit cycle by a single oscillator.  
  
  \item A procedure to learn the sensor model  is presented.  This is in contrast to~\cite{taghvaei2014coupled,tixian2019control}, where the sensor model is assumed to be known. 
  
\item The learning framework is numerically demonstrated in a
  simulation environment.  The main innovation is the
  parametrization of the Q-function (or the Hamiltonian).  
\end{enumerate}

Taken together, the numerical results of this paper demonstrate an
end-to-end architecture for sensori-motor control of bio-locomotion.   
These results are likely to spur comparative studies as well as
theoretical investigations of learning in bio-locomotion.  



The remainder of this paper is organized as follows: 
The snake system problem is formulated in Section~\ref{sec:ProbForm}. 
The control problem solution is described in Section~\ref{sec:SolAppr}.
The numerical results of the snake system appear in Section~\ref{sec:Numerics}. 

\section{Problem Formulation}
\label{sec:ProbForm}

\subsection{Modeling}
\begin{figure}[t]
	\centering
	\includegraphics[width=0.9\columnwidth]{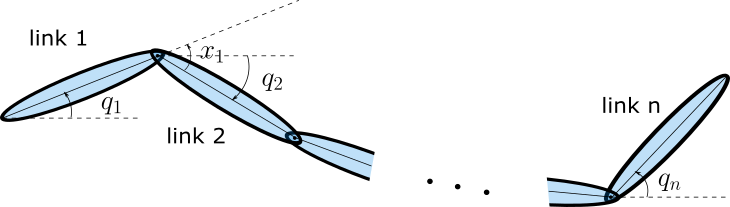}
	\caption{Schematic of the $n$-link system for the snake robot.}
	\label{fig:snake_sys}
\end{figure}

The model of snake robot, described next, closely follows~\cite{Iwasaki-snake}. Consider a system of  $n$ planar rigid links, connected by single degree of freedom  joints as depicted in Fig.~\ref{fig:snake_sys}. The system is placed on a horizontal surface, subject to friction. 
The $j$-th joint is equipped with torque actuator (motor) with drive torque $\tau_j$, linear torsional spring with coefficient $\kappa_j$, and viscous friction with coefficient $\zeta_j$ for $j=1,\ldots,n-1$. 
It is assumed that each link has uniformly distributed mass. For link $j$, $m_j$ denotes its mass, $l_j$ denotes its half length, and $J_j$ denotes its moment of inertia about the center of mass.

The absolute orientation of $j$-th link, with respect to a global inertial frame, is denoted by $q_j \in [0,2\pi]$, and the position of the center of mass is denoted by $r\cm \in \Re^2$. As a result, $(q,r\cm) \in [0,2\pi]^n \times \Re^2$, with $q:=(q_1,\ldots,q_n)$, represents the configuration of the $n$-link system. 

The configuration  is divided into two sets: (i) the shape variable; (ii) and the group variable. The shape variable, $x=(x_1,\ldots,x_{n-1}) \in [0,2\pi]^{n-1}$, are the relative angle between the links, defined as $x_j=q_j-q_{j+1}$ for  $j=1,\ldots,n-1$. 
%
The group variable $(\psi,r\cm) \in SE(2)$ comprises the global orientation of the system,
\begin{equation}
\psi:=\frac{1}{n}\sum_{j=1}^{n} q_j
\label{eq:q-to-psi}
\end{equation}
and the position of center of mass $r\cm$. The group variable is an element of the group of planar rigid body motions $SE(2)$. 
  
An open loop periodic input is assumed for torque actuators,
\begin{equation}
\tau_j(t)= {\tau_0}_j\sin(\omega_0t + \beta_j),\quad \text{for}\ j=1,\ldots,n-1
\label{eq:open-loop-torque}
\end{equation} 
where $\omega_0$ is frequency, ${\tau_0}_j$ is the amplitude, and $\beta_j$ is the phase. The particular form of the periodic input is not important. 
For the purpose of numerics, this input is chosen to induce a nominal gait, which leads to forward motion.

The friction force exerted at each link comprises of three component: A force component directed normal to the link, a force component directed tangent to the link, and a torque. The models for these components are,
\begin{equation*}
\begin{aligned}
\text{normal friction force}&=-c_{n,j}m_j v_{j}\cdot \hat{n}_j \\
\text{tangent friction force}&=-c_{t,j}m_j v_{j}\cdot \hat{t}_j \\
\text{friction torque}&= -c_{n,j}J_j\dot{q}_j
\end{aligned}
\end{equation*}where 
$\hat{n}_j$, $\hat{t}_j$ are the normal and tangent unit vectors to link $j$, $v_j$ is the velocity of link $j$, and $c_{t,j},c_{n,j}$ are the friction coefficients in the tangent and normal directions, respectively.
For snake robot, these coefficients are different ($c\NN\gg c\TT$) which is believed be essential for forward locomotion~\cite{Hirose}.

For the snake robot model problem, the control input enters via change in friction coefficients as,
\begin{equation}
c_{n,j}(t)=\bar{c}_{n,j}(1+u_j(t)),\quad \text{for}\ j=1,\ldots,n
\label{eq:u-def}
\end{equation}
where $\bar{c}_{n,j}$ is the nominal friction coefficient normal to
link $j$, and $u_j(t)$ represents a small time-dependent perturbation
due to control. 

\subsection{Dynamics}
The dynamics of the system is given by a second order ode for the shape variable, and a first-order ode for the group variable: 
\begin{align}
\ddot{x}(t) &= \tilde{F}_x(x(t),\dot{x}(t),\tau(t),u(t)),
\label{eq:i-dynamics-u}\\
\frac{\ud}{\ud t}
\begin{bmatrix}
\psi(t) \\ r\cm(t)
\end{bmatrix} &= \begin{bmatrix}
 \tilde{F}_\psi(x(t),\dot{x}(t),u(t)) \\  \tilde{F}_r(x(t),\dot{x}(t),u(t))
\end{bmatrix}  
\label{eq:g-dynamics-u}
\end{align}
The derivation of the dynamic equations and the explicit form of the functions $\tilde{F}_x$, $\tilde{F}_\psi$, and $\tilde{F}_r$ appears in Appendix~\ref{apdx:dynamics}. In this paper, the explicit form of these functions are assumed to be unknown. 

\subsection{Observation process}
\label{subsec:obsv}
The shape variable and its velocity $(x,\dot{x})$ are 
assumed not to be fully observed . To estimate $(x,\dot{x})$, each joint is equipped with a sensor that provides noisy measurements of the shape variable and its velocity. The model for the sensor at the $j$-th joint is 
\begin{equation}
	\ud Z_j(t) = \tilde{h}_j(x_j(t),\dot{x}_j(t))\ud t + \sigma_W \ud W_j(t),\quad \text{for} j=1,\ldots,n-1
	\label{eq:obsv}
\end{equation}
where $W_j(t)$ is a standard Wiener process and $\sigma_W$ is the standard 
deviation parameter.  
The explicit form of the function $\tilde{h}_j(\cdot)$ in the observation model is assumed to be unknown. However, it is assumed that $\tilde{h}_j(\cdot)$ is only a function of $(x_j,\dot{x}_j)$.

\subsection{Optimal control problem}
\label{subsec:CtrlProb}


%

 The control objective is to find a control input $u(t)$ that turns the robot, while the robot
is moving forward with a nominal gait produced by the uncontrolled open-loop input torque $\tau(t)$ according to~\eqref{eq:open-loop-torque}. 
The control objective is modeled as a discounted infinite-horizon optimal
control problem:
\begin{equation}
    \tilde{J}(x(0),\dot{x}(0)) 
    = \underset{u(\cdot)}{\min}\ {\sf E} \left[ 
    \int_0^\infty e^{-\gamma t} \tilde{c}(x(t),\dot{x}(t),u(t)) \ud t \right]
    \label{eq:opt-cont}
\end{equation}
subject to the dynamic constraints~\eqref{eq:i-dynamics-u}.
Here, $\gamma>0$ is the discount rate and the cost function 
\begin{equation}
\tilde{c}(x,\dot{x},u)=\tilde{F}_\psi(x,\dot{x},u)+\frac{1}{2\epsilon}\|u\|_2^2
\label{eq:cost}
\end{equation}
where $\tilde{F}_\psi(x,\dot{x},u)=\dot{\psi}$ is the rate of change
of the global orientation $\psi$,  $\|u\|_2^2=\sum_{j=1}^n u_j^2$, and
$\epsilon>0$ is the control penalty parameter.  
The minimum is over all control inputs $u(\cdot)$ adapted to the filtration 
$\mathcal{Z}_t:=\sigma(Z(s);s\in[0,t])$ generated by the observation process. 

The cost function is chosen so that, minimizing the cost
leads to negative net change in the global orientation $\psi$, 
which corresponds to the clockwise rotation.    

\section{Solution Approach}
\label{sec:SolAppr}
Solving the optimal control problem~\eqref{eq:opt-cont} 
is challenging because:
\begin{enumerate}
	\item The function $\tilde{F}_x$ in the dynamics model~\eqref{eq:i-dynamics-u} for the $(n-1)$-dimensional shape variable $x$ is assumed to be unknown.
		 The model is highly nonlinear 
		due to the details of the geometry and contact forces with 
		the environment (see~\eqref{eq:i-dynamics} in Appendix~\ref{apdx:dynamics}). 
	\item The explicit form of the function $\tilde{F}_\psi$ that appears in the cost function~\eqref{eq:cost}
		is assumed to be unknown.
	\item The shape variable $(x,\dot{x})$ is not fully observed. 
	\item The explicit form of the observation functions $\tilde{h}_j(\cdot)$ 
		in~\eqref{eq:obsv} are assumed to be unknown. 
          
\end{enumerate}

The following steps are used to overcome these challenges:

\subsection{Step 1. Phase modeling}
\label{subsec:model-reduction}

\begin{figure}[t]
	\centering
	\includegraphics[width=0.9\columnwidth]{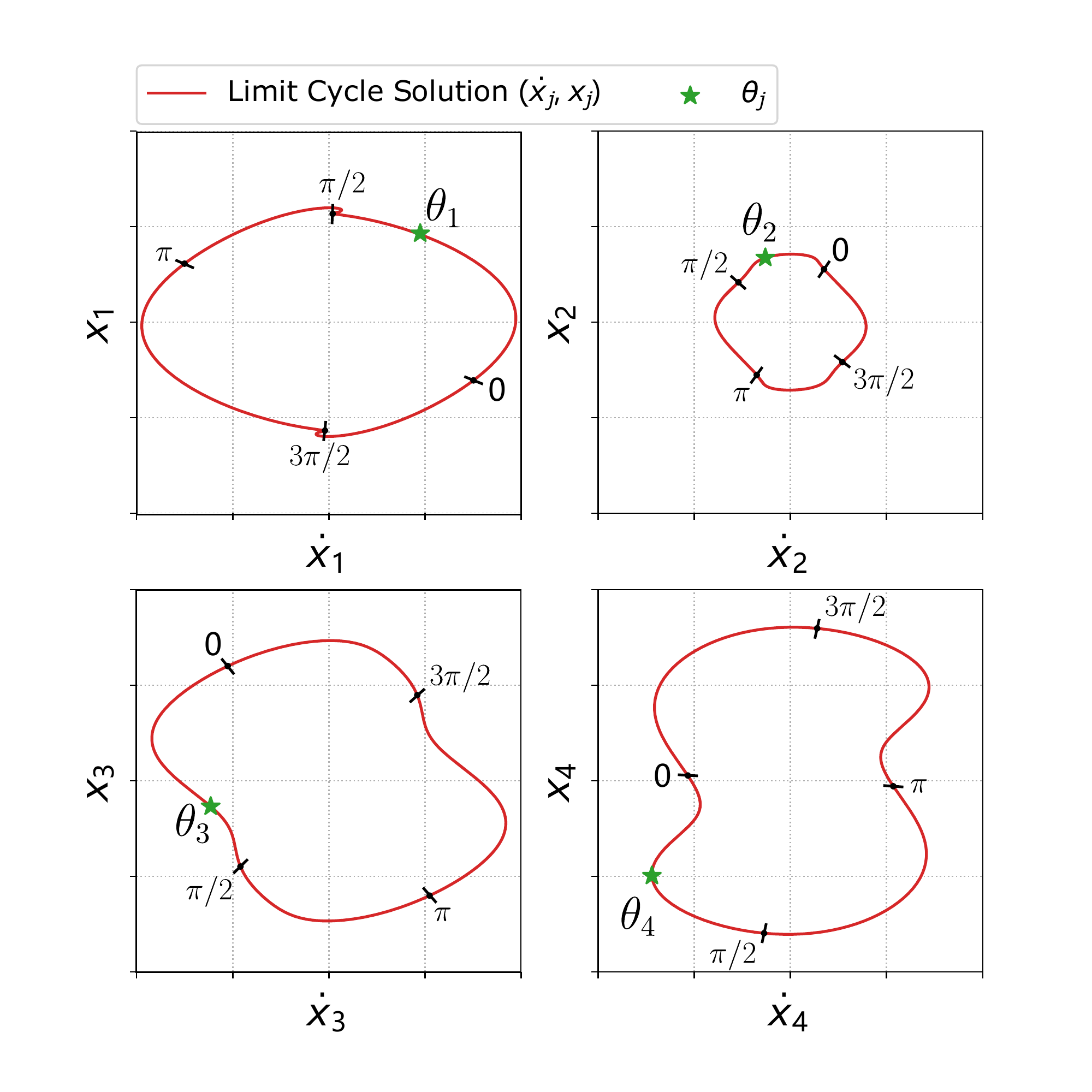}
	\caption{The limit cycle solution for the shape variable
		$(x(t),\dot{x}(t))$ under the periodic torque input~\eqref{eq:open-loop-torque},  for a $5$-link system. Each limit cycle is parametrized  with a phase variable $\theta_j \in [0,2\pi]$. 
		}
	\label{fig:limit-cycle}
\end{figure}

Consider the second-order differential equation \eqref{eq:i-dynamics-u} 
for the shape variable $x$ under the open-loop periodic input $\tau(t)$ 
in~\eqref{eq:open-loop-torque}. The following assumption is made
concerning its solution:

\begin{romannum}
	\item[\textbf{\textit{Assumption~A1}}] 
	Under periodic forcing $\tau(t)$ as in \eqref{eq:open-loop-torque}, 
	the solution $(x_j(t),\dot{x}_j(t))$ to \eqref{eq:i-dynamics-u} is an 
	isolated asymptotically stable periodic orbit (limit cycle) 
	with period ${2\pi}/{\omega_0}$ 
 for $j=1,\ldots,n-1$ (see Figure~\ref{fig:limit-cycle}).
\end{romannum}

Denote the set of points on the limit cycle of $(x_j(t),\dot{x}_j(t))$ as $\mathcal{P}_j\subset\mathbb{R}^{2}$. 
Each limit cycle solution is parameterized by a phase coordinate 
$\theta_j\in[0,2\pi)$ in the sense that there exists an invertible map 
$\xlc_j:[0,2\pi) \to \mathcal{P}_j$ such that 
$\xlc_j(\theta_j(t))=(x_j(t),\dot{x}_j(t))$, 
where $\theta_j(t)=(\omega_0 t+\theta_j(0))$ mod $2\pi$, for $j=1,\ldots,n-1$.  
The definition of the phase variable is extended locally in a small neighborhood of
the limit cycle by using the notion of isochrons~\cite{izhikevich2007dynamical}.

Let $\thetan(t):=(\theta_{1}(t),\ldots,\theta_{n-1}(t))$ denote the vector 
of all the phase variables, and $\xlc(\theta):=(\xlc_1(\theta_1),\ldots,\xlc_{n-1}(\theta_{n-1}))$. In terms of $\theta(t)$, the first-order dynamics of the 
group variable in~\eqref{eq:g-dynamics-u} is expressed as
\begin{equation}
	\frac{\ud}{\ud t}
	\begin{bmatrix}
		\psi(t) \\ r\cm(t)
	\end{bmatrix} = \begin{bmatrix}
 \tilde{F}_\psi(\xlc(\theta(t)),u(t))\\  \tilde{F}_r(\xlc(\theta(t)),u(t))
	\end{bmatrix}  =:\begin{bmatrix}
	F_\psi(\theta(t),u(t)) \\  F_r(\theta(t),u(t))
	\end{bmatrix}  
	\label{eq:g-dynamics-theta}
\end{equation}
and the observation model \eqref{eq:obsv} is 
\begin{equation}
	\ud Z_j(t) = h_j(\theta_j(t))\ud t + \sigma_W \ud W_j(t),\quad \text{for}\ j=1,\ldots,n-1
    \label{eq:obsv_theta}
\end{equation}
where 
$h_j(\theta_j) := \tilde h_j(\xlc_j(\theta_j))$. 

The optimal control problem \eqref{eq:opt-cont} 
in terms of the phase vector is given by
\begin{equation}
    J(\theta(0)) 
    = \underset{u(\cdot)}{\min}\ {\sf E} \left[ 
    \int_0^\infty e^{-\gamma t} c(\thetan(t),u(t)) \ud t \right]
    \label{eq:opt-cont-theta}
\end{equation}
where $c(\theta,u)=F_\psi(\theta,u)+\frac{1}{2\epsilon} \|u\|_2^2$ and the minimum is 
over all control inputs $u(\cdot)$ adapted to the filtration
$\mathcal{Z}_t$. 
%


The new problem is described by a single phase vector $\thetan$ 
instead of coupled shape variables $x$ and $\dot{x}$. 
With $u(t)\equiv 0$, the dynamics is described by the oscillator model 
$\thetan_j(t)=(\omega_0 t+\theta_j(0))$ 
mod $2\pi$ for $j=1,\ldots,n-1$. 
	Now, in the presence of (small) control input, the
	dynamics need to be augmented by an additional term $\epsilon g(\theta,u)$ due to control:
	\begin{equation}\label{eq:controlled_theta}
	\ud \theta (t) = (\omega_0 {1}_{n-1} + \epsilon g(\theta(t),u(t))) \ud t
	\end{equation}
where $1_{n-1}=[1,\ldots,1]^T \in \Re^{n-1}$. 
 



\subsection{Step 2. Learning observation model}
\label{subsec:obsv-model}

The explicit form of the function $h_j(\cdot)$ in the  observation model~\eqref{eq:obsv_theta} is not known. 
It is approximated using a linear combination of the Fourier basis functions:
\begin{equation}
h_j(\theta_j)\approx h_j(\theta_j;r_j) :=r_j^T \phi_h(\theta_j),\quad \text{for}\ j=1,\ldots,n-1
\label{eq:obsv_model}
\end{equation}
where $\phi_h$ is a vector of $M_h$ Fourier basis functions (e.g $\phi_h(\vartheta)=(\sin(\vartheta),\cos(\vartheta))$), and  $r_j\in\mathbb{R}^{M_h}$ is a vector of $M_h$ weights. The weights are initialized at zero and updated in an online fashion according to   
\begin{equation}
\ud r_j(t) = \alpha_h(t)\left[\ud Z_j(t)-\hat{h}_j(t)\ud t\right] \Expect[\phi_h(\theta_j(t))|\clZ_t]
\label{eq:r-update}
\end{equation}
where $\alpha_h(t)$ is the learning rate, and $\hat{h}_j(t):=\Expect[h_j(\theta_j(t),r_j(t))|\clZ_t]$.  In numerical implementation, the conditional expectations are approximated using the feedback particle filter, described next. 


\subsection{Step 3. Feedback particle filter (FPF)}
\label{subsec:FPF}

The feedback particle filter algorithm is used  to obtain the posterior 
distribution of the phase vector $\thetan(t)$, governed by dynamics~\eqref{eq:controlled_theta}, given the noisy observations 
\eqref{eq:obsv_theta}. 
The filter comprises $N$ stochastic processes 
$\{\theta^{i}(t):1\leq i\leq N\}$, where $\theta^{i}(t)\in[0,2\pi]^{n-1}$ 
is the state of the $i$-th particle (oscillator) at time $t$. 
The particles evolve according to
\begin{align}
    \ud &\thetan^i(t)  = \omega^i 1_{n-1} \ud t + \epsilon g(\theta^i(t),u(t))
	\ud t \nonumber \\ 
	& + \sum_{j=1}^{n-1}\frac{\v_j(\theta^i(t),t)}{\sigma_W^2} \circ 
    \left(\ud Z_j(t) - \frac{h_j(\theta_j^i(t),r_j(t)) + \hat{h}_j(t)}{2} \ud t \right)
    \label{eq:g-dynamicsPF}
\end{align}
where $\omega^i\sim\text{Unif}([\omega_0-\delta,\omega_0+\delta]^{n-1})$ is the 
frequency of the $i$-th oscillator, 
$\hat{h}_j(t):={\sf E}[h_j(\theta_j(t),r_j(t))|\mathcal{Z}_t]$, and the notation $\circ$ denotes Stratonovich integration. In numerical implementation $\hat{h}_j(t) \approx \frac{1}{N} 
\sum_{i=1}^N h_j(\theta_j^i(t),r_j(t))$.  

The algorithm involves $n-1$ gain functions $\v_j(\theta,t)$ for $j=1,\ldots,n-1$, where the $j$-th gain function corresponds to the $j$-th observation signal. Each gain function is a $(n-1)$-dimensional vector expressed as $\v_j(\theta,t)=(\v_{j,1}(\theta,t),\ldots,\v_{j,n-1}(\theta,t))\in \Re^{n-1}$. The gain function is the solution of a certain partial differential equation. In practice, the gain function is numerically approximated using the Galerkin algorithm. The details of the Galerkin algorithm appears in~\cite{tilton2013multi}. 

Given the particles, the conditional expectation $\Expect[f(\theta(t))|\clZ_t]$ of a given function $f(\cdot)$ is approximated as $ \frac{1}{N}\sum_{i=1}^N f(\theta^i(t)) $.



	
%

	\medskip
	\begin{remark}
	There are two manners in which control input $u(t)$ affects the
	dynamics of the filter state $\thetan^i(t)$:
	\begin{enumerate}
	\item The $O(\epsilon)$ term $\epsilon g(\cdot,u(t))$ which models the effect
	of dynamics; 
	\item The FPF update term which models the effect of sensor
	measurements.  This is because the control input $u(t)$ affects the
	state $(x(t),\dot{x}(t))$ (see~\eqref{eq:i-dynamics-u}) which in turn affects the
	sensor measurements $Z(t)$ (see~\eqref{eq:obsv}).  
	\end{enumerate}
	\end{remark}


\subsection{Step 4. Q-learning}
\label{subsec:Q-learning}
%

With the constructed FPF, we can now express the partially observed 
optimal control problem~\eqref{eq:opt-cont-theta} as a fully observed 
optimal control problem in terms of oscillator states 
$\thetaN(t)= (\thetan^1(t),\ldots,\thetan^N(t))$ according to
\begin{equation}
    \JN(\thetaN(0)) 
    = \underset{u(\cdot)}{\min}\ {\sf E} \left[ 
    \int_0^\infty e^{-\gamma t} \cN(\thetaN(t),u(t)) \ud t \right]
    \label{eq:opt-cont-theta-i}\end{equation}
subject to~\eqref{eq:r-update}-\eqref{eq:g-dynamicsPF}, where the cost $\cN(\thetaN,u) := \frac{1}{N}
\sum_{i=1}^N c(\thetan^i,u)$ and the minimization is over all control laws 
adapted to the filtration $\mathcal{X}_t:=\{\thetan^i(s); ~s\leq t, 
1\leq i\leq N\}$. The problem is now fully observed because the states of 
oscillators $\thetaN(t)$ are known. This approach closely follows~\cite{mehta2013feedback}.


	The analogue of the Q-function for continuous-time systems is 
	the Hamiltonian function:
	\begin{equation}\label{eq:Hamiltonian}
		\HN(\thetaN,u) = \cN(\thetaN,u) + \mathcal{D}_u \JN(\thetaN)
	\end{equation}
	where  $\mathcal{D}_u$ is the generator for~\eqref{eq:g-dynamicsPF} defined such that 
	$\frac{\ud}{\ud t}\Expect[\JN(\thetaN(t))]=\mathcal{D}_u\JN(\thetaN(t))$.  

	The dynamic programming principle for the discounted problem implies:
	\begin{equation}\label{eq:dynamic-programming}
		\min_u~ \HN(\thetaN,u) = \gamma \JN(\thetaN)
	\end{equation}
	Substituting this into the definition of the Hamiltonian~\eqref{eq:Hamiltonian} 
	yields the fixed-point equation:
	\begin{equation}
		\mathcal{D}_u~\HminN(\thetaN) = -\gamma (\cN(\thetaN,u) - \HN(\thetaN,u))
		\label{eq:g-dynamicsixed-pt}
	\end{equation}
	where $\HminN(\thetaN):=\min_u~\HN(\thetaN,u)$. 
	This is equivalent to the fixed-point equation that appears
	in the Q-learning algorithm in discrete-time setting. 

\medskip
\noindent
\textbf{Linear function approximation:}
The Hamiltonian function is 
approximated as the linear combination of $M$ real-valued basis 
functions $\{\phi_m(\theta,u)\}_{m=1}^M$ as follows:
\begin{equation}
    \begin{aligned}
        \hatHN(\thetaN,u;w) & = \frac{1}{N} \sum_{i=1}^N w^T\phi(\thetan^i,u) 
    \end{aligned}
    \label{eq:LFA}
\end{equation}
where $w\in \Re^M$ is a vector of weights and $\phi = (\phi_1,\ldots,\phi_M)^T$ 
is a vector of basis functions. 
Thus, the infinite-dimensional problem of learning the Hamiltonian function 
is reduced to the problem of learning the $M$-dimensional weight
vector $w$.

We define the point-wise Bellman error as follows:
\begin{equation}
    \begin{aligned}
        \mathcal{E}(\thetaN,u;w):=&\mathcal{D}_u\hatHminN(\thetaN;w) 
        \\&+ \gamma (\cN(\thetaN,u) - \hatHN(\thetaN,u;w))
    \end{aligned}
    \label{eq:Bellman-error}
\end{equation} 
where $\hatHminN(\thetaN;w):= \min_u~\hatHN(\thetaN,u;w)$. 

Then a gradient descent algorithm to learn the weights is:
\begin{equation}
    \frac{\ud}{\ud t} w(t) 
    = -\frac{1}{2}\alpha(t) \nabla_w \mathcal{E}^2(\thetaN(t),u(t);w(t))
    \label{eq:update-w}
\end{equation}
where $\alpha(t)$ is the learning rate  and $u(t)$ 
is chosen to explore the state-action space. For the  convergence analysis of the
Q-learning algorithm, see~\cite{szepesvari1998asymptotic,moulines2011non}.

\medskip
Given a learned weight vector $w^*$, the learned optimal control policy is given by:
\begin{equation}\label{eq:learned-u}
\hat{u}^*(\thetaN;w^*) = \underset{v}{\arg\min}\ 
\hatHN(\thetaN,v;w^*) 
\end{equation}

\subsection{Information structure}
In order to implement the FPF algorithm~\eqref{eq:g-dynamicsPF}, it is necessary to know the model for $g(\thetan,u)$. The function $g(\thetan,u)$ represents the effect of the control input on the limit cycle. 
However, it is numerically observed that the control input has negligible effect on the limit cycle solution. 
Thus, in the simulation results presented next, the term 
$\epsilon g(\theta^{i}(t),u(t))$ is ignored.  


In the Q-learning algorithm, the generator $\mathcal{D}_u$ is approximated numerically as 
\[
\mathcal{D}_u\hatHminN(\thetaN(t)) \approx \frac{\hatHminN(\thetaN(t+\Delta t)) 
	- \hatHminN(\thetaN(t)) }{\Delta t}
\] 
where $\Delta t$ is the discrete time 
step-size and $\{\theta^i(t)\}_{i=1}^N$ is the state of the oscillators at time $t
$. 

The function $F_\psi(\theta(t),u(t))$ that appears in the cost function is numerically approximated as
	\begin{equation*}
F_\psi(\thetan(t),u(t))=\dot{\psi}(t)\approx \frac{\psi(t+\Delta t) 
	- \psi(t)}{\Delta t}
\end{equation*}
where $\Delta t$ is the discrete time 
step-size in the numerical algorithm and $\psi(t)$ is available through a
(black-box) simulator, that simulates the dynamics~\eqref{eq:i-dynamics-u} and~\eqref{eq:g-dynamics-u}.




\subsection{Distributed aspect of the architecture }
The FPF algorithm~\eqref{eq:g-dynamicsPF} is simplified to $n-1$
independent filters as follows.  By ignoring the $\epsilon
g(\theta,u)$ term in~\eqref{eq:controlled_theta}, the evolution of the
each component $\theta_j \in [0,2\pi]$ of the $n-1$ dimensional phase
variable $\theta\in[0,2\pi]^{n-1}$ becomes independent of each
other. Moreover, the observation functions $h_j(\cdot)$ for
$j=1,\ldots, n-1$ in the sensor model~\eqref{eq:obsv_theta} are
independent of each other, in the sense that $h_j(\cdot)$ is a function of only $\theta_j$. Therefore, the posterior distribution of the phase variable $\theta$ is the product of $n-1$ independent distributions for $\theta_j$. With independent posterior distribution, the $j$-th gain function $\v_j(\theta,t) \in \Re^{n-1}$ in the FPF algorithm~\eqref{eq:g-dynamicsPF} takes the form $\v_j(\theta,t)=(0,\ldots,0,\v_{j,j}(\theta_j,t),0,\ldots,0)\in \Re^{n-1}$. As a result, the FPF algorithm is decomposed to $n-1$ independent filters.  The evolution of particles $\{\theta^i_j(t)\}_{i=1}^N$ for the $j$-th filter is
\begin{align}
\ud &\theta^i_j(t)  = \omega^i \ud t \nonumber \\ 
& + \frac{\v_{j,j}(\theta^i_j(t),t)}{\sigma_W^2} \circ 
\left(\ud Z_j(t) - \frac{h_j(\theta_j^i(t),r_j(t)) + \hat{h}_j(t)}{2} \ud t \right)
\label{eq:g-dynamicsPF-j}
\end{align}
Therefore, the FPF algorithm for each joint is simulated independently from the other FPFs for other joints,  in a distributed manner as shown in Figure~\ref{fig:ctrl_arch}.    

The learned control input is also designed to take distributed structure, in the sense that the control input to each  link depends only on the phase variable of its adjacent joints. The distributed structure is enforced by a careful selection of basis functions for the Hamiltonian in~\eqref{eq:LFA}.  The selected basis functions consist of three groups:
\begin{equation}
\begin{aligned}
	\text{group 1:}&\quad \{\Phi(\theta_j)\}_{j=1}^{n-1} \\
	\text{group 2:}&\quad \{u_j\Phi(\theta_j),u_{j+1} \Phi(\theta_{j})\}_{j=1}^{n-1} \\
	\text{group 3:}&\quad \{\frac{1}{2}u_j^2\}_{j=1}^{n} 
\end{aligned}
\label{eq:basis-functions}
\end{equation}
where $\Phi(\vartheta)=(\Phi_1(\vartheta),\ldots,\Phi_{M_F}(\vartheta))$ is a vector of selected Fourier basis functions (e.g $\Phi(\vartheta)=(\sin(\vartheta),\cos(\vartheta))$). With this particular form of basis functions, the $j$-th component of the learned control input~\eqref{eq:learned-u} takes the following form: 
\begin{equation}
\hat{u}^*_j(\thetaN,w^*) = \frac{1}{N}\sum_{i=1}^N\sum_{m=1}^{M_F} a_{j,m}\Phi_m(\theta^i_j) + b_{j,m}\Phi_m(\theta^i_{j-1})
\label{eq:u-distributed}
\end{equation}
where the constants $a_{j,m}$ and $b_{j,m}$ depend on the value of optimal weight vector $w^*$, and the convention $b_{1,m}=a_{n,m}=0$ is assumed, for $m=1,\ldots,M_F$. According to the formula~\eqref{eq:u-distributed}, the  control input to $j$-th link, only depends on the phase of the adjacent joints $\theta_j$ and $\theta_{j-1}$.  The overall numerical procedure is summarized in Algorithm~\ref{algm:QL-OptControl}.

\begin{algorithm}[t]
	\caption{The proposed numerical algorithm}
	\label{algm:QL-OptControl}
	\begin{algorithmic}[1]
		\Require  A simulator for~\eqref{eq:i-dynamics-u}-\eqref{eq:g-dynamics-u}-\eqref{eq:obsv}.
		\Ensure Optimal control policy $\hat{u}^*(\thetaN;w)$.
		\State Initialize weight vector $w_0$
		\State Initialize particles $\Big\{\{\theta^i_{j}(0)\}_{i=1}^N
		\Big\}_{j=1}^{n-1}\sim\text{Unif}([0,2\pi])$;
		\For{the $k=1$ to $n_T\frac{2\pi}{\omega_0 \Delta t} $}
		\State Choose control input $u(k)$ according to \eqref{eq:u_sin};
		\State Input $u(k)$ to simulator and output $Z(k),\psi(k)$;
			\For{the $j=1$ to $n-1$}
\State Compute $\hat{h}_j(k)=\frac{1}{N}\sum_{i=1}^N h_j(\theta_j^i(k),r_j(k))$  and $\Delta Z_j(k) = Z_j(k+1) - Z_j(k)$
		\State Update the weights for observation model
		\begin{align*}
			r_{j}& (k+1)= r_{j}(k) + \alpha_{h}\left[\Delta Z_j(k) -\hat{h}_{j}(k)\Delta t\right]\frac{1}{N}\sum_{i=1}^N \phi_h(\theta_{j}(k)^i)
		\end{align*}
		\State Update the particles 
		\begin{equation*}
			\begin{aligned}
				\theta_{j}^i&(k+1)  = \theta_{j}^i(k) + \omega^i \Delta t \\
				& + \frac{\v_{j,j}(\theta_{j}^i(k),k)}{\sigma_W^2}  
				(\Delta Z_j(k)- \frac{h_j(\theta_{j}^i(k),r_j(k))+\hat{h}_j(k)}{2}
				\Delta t )
			\end{aligned}
		\end{equation*}
		\EndFor
		\State Compute cost $c(k)= \frac{\psi(k+1)-\psi(k)}{\Delta t} + \frac{1}{2\epsilon}\|u(k)\|_2^2$ 
		\State Compute Bellman error 
		\begin{equation*}
		\mathcal{E}(k)=\mathcal{D}_u \hatHminN(k) 
		+ \gamma(c(k) - \hatHN(\thetaN(k),u(k);w(k)))
		\end{equation*}
		\State Update weight 
		$w(k+1)=w(k)-\Delta t \alpha\mathcal{E}(k)\nabla_w\mathcal{E}(k)$
		\EndFor
		\State Output the learned control $\hat{u}^*(\thetaN;w(k))$ 
		from \eqref{eq:learned-u}.
	\end{algorithmic}
\end{algorithm}



\addtolength{\textheight}{-1cm}   

\section{Numerics}
\label{sec:Numerics}


The following  numerical results are for the snake robot with $n=5$ links.  The numerical results are based on Algorithm~\ref{algm:QL-OptControl}.
The simulation parameters are tabulated in Table~\ref{tab:num_para}.

\begin{table}[tb]
	\centering
	\caption{Parameters for Numerical Simulation}
	\begin{tabular}{cccc}
		\hline
		\hline\noalign{\smallskip}
		Parameter & Description & \multicolumn{2}{c}{Numerical value} \\
		\hline
		\hline\noalign{\smallskip}
		\multicolumn{3}{c}{{\bf Sensor \& FPF}}\\
		$\Delta t$ & Discrete time step-size & $0.02$ \\
        $\sigma_W$ & Noise process std. dev. & $0.1$ \\
		$N$ & Number of particles & $100$ \\
		$\delta$ & Heterogeneous parameter & $0.05$ \\
		\hline
		\hline\noalign{\smallskip}
		\multicolumn{3}{c}{{\bf Q-learning}}\\
		$n_e$ & number of episodes & $200$ \\
		$n_T$ & number of periods in each episode & $10$ \\ 
		$\epsilon$ & Control penalty parameter & $10.0$ \\
		$\gamma$ & Discount rate & $0.50$ \\
		$\alpha$ & Learning gain for Q-learning & $0.01$ \\
		$\alpha_h$ & Learning gain for observation model & $0.01$ \\
        \hline 
	\end{tabular}
	\label{tab:num_para}
\end{table}

%

\begin{figure}[t]
	\centering
	\includegraphics[width=0.9\columnwidth]{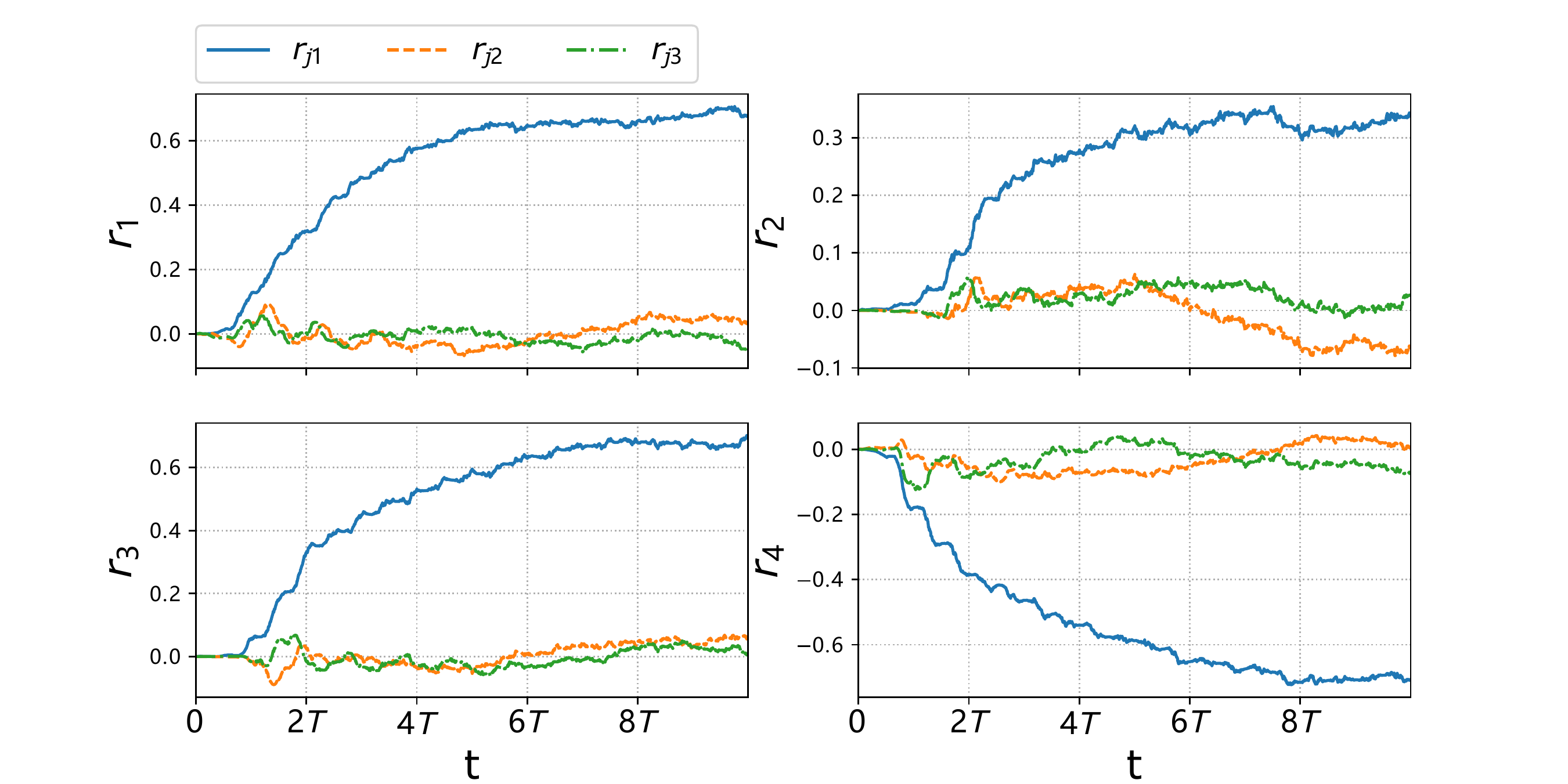}
	\caption{The time-trace of the weights for learning the observation model according to~\eqref{eq:r-update}.}
	\label{fig:learning-r}
\end{figure}

\begin{figure}[t]
	\centering
	\includegraphics[width=0.9\columnwidth,keepaspectratio=true]
	{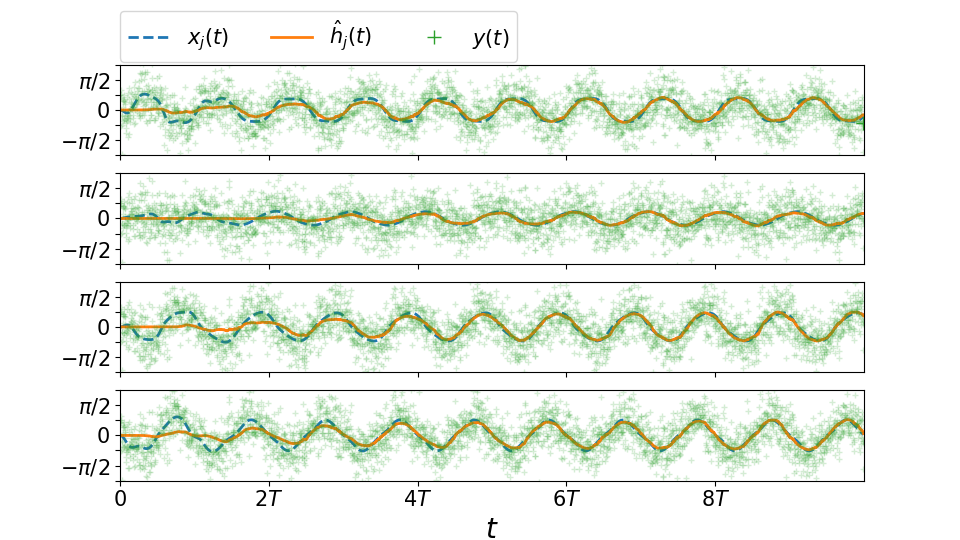}
	\caption{The figure contains three signals: (i) $y_j(t)$: the noisy observation signal from the observation model~\eqref{eq:obsv} where $y(t)=(Z(t+\Delta t)-Z(t))/\Delta t$; (ii) The exact observation signal  $\tilde{h}_j(x_j(t),\dot{x}_j(t))=x_j(t)$; (iii) and the approximation $\hat{h}_j(t)=N^{-1}\sum_{i=1}^N h_j(\theta_j^i(t),r_j(t))$.}
	\label{fig:obsv}
\end{figure}


\subsection{Learning the observation model and FPF}
\label{subsec:Num-PM-FPF}
The observation signal $y_j(t):=(Z_j(t+\Delta t)-Z_j(t))/\Delta t$, for $j=1,2,3,4$ is depicted in Figure~\ref{fig:obsv}.
The signal is generated according to~\eqref{eq:obsv}, with observation function  taken as $\tilde{h}_j(x_j,\dot x_j)=x_j$.  The noise strength 
$\sigma_w = 0.1$.

The Fourier basis functions used to approximate the observation function according to~\eqref{eq:obsv_model} are 
\begin{equation*}
\phi_h(\vartheta) = (\sin(\vartheta),\sin(2\vartheta),\cos(2\vartheta))
\end{equation*}
Including the $\cos(\vartheta)$ in the basis functions is redundant
because of the degeneracy in defining the phase. 

\begin{figure}[t]
	\centering
	\includegraphics[width=0.9\columnwidth,keepaspectratio=true]
	{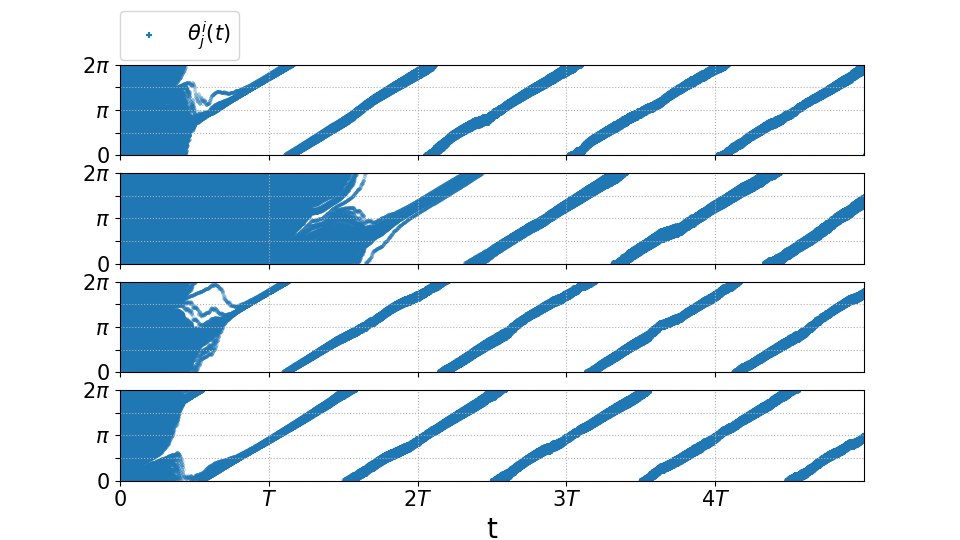}
	\caption{Time trace of $N=100$ particles for  the four independent FPF algorithm~\eqref{eq:g-dynamicsPF-j}. The empirical distribution of the particles for the $j$-th FPF approximates the posterior distribution of the phase variable corresponds to the $j$-th joint of the $5$-link system.}
	\label{fig:multi-FPF}
\end{figure}

The gradient descent algorithm~\eqref{eq:r-update}, to learn the weights $r_j$, and the FPF algorithm~\eqref{eq:g-dynamicsPF-j}, for the $j$-th joint are simulated, for $j=1,2,3,4$. The time-trace of the weights $r_j(t)$, and the trajectory of particles $\theta^i_j(t)$, are depicted in Figure~\ref{fig:learning-r} and~\ref{fig:multi-FPF} respectively. 

The performance of the observation model learning algorithm and the FPF algorithm is observed in Figure~\ref{fig:obsv}. The figure includes three signals: (i) The noisy measurements $y_j(t)$; (ii) the value $\tilde{h}_j(x_j(t),\dot{x}_j(t))=x_j(t)$; (iii) and the approximation $\hat{h}_j(t)=\frac{1}{N}\sum_{i=1}^N h_j(\theta_j^i(t),r_j(t))$, which involve the learned wights $r_j(t)$ and the particles $\{\theta_j^i(t)\}_{i=1}^N$.  It is observed that the approximation $\hat{h}_j(t)$ converges to the exact value   $h_j(x_j(t),\dot{x}_j(t))$,  as the learning for the weights  converge  and particles become synchronized.

\subsection{Q-learning}
\label{subsec:Num-QL}

The Q-learning algorithm is simulated for $200$ episodes. Each episode starts with random initialization of the state, and continues for $n_T=10$ periods. 

The basis function used to approximate the Hamiltonian in~\eqref{eq:LFA}  
are chosen according to~\eqref{eq:basis-functions} 
with $\Phi(\vartheta)=(\cos(\vartheta),\sin(\vartheta),\cos(2\vartheta),
\sin(2\vartheta))$. 


The weights for the $\frac{1}{2}u_j^2$ basis function  are initialized randomly  with
uniform distribution $\text{Unif}([0.09,0.11])$.
The rest of the weights are initialized according to $\text{Unif}([-0.1,0.1])$

For the purpose of exploration, the control input $u(t)$ to be used in~\eqref{eq:update-w} is chosen as a combination of sinusoidal functions 
with irrational frequencies as follows:
\begin{equation}
u_j(t) = A\sin(\sqrt{2}\omega_0t+ \frac{j\pi}{5}) + A\sin(\pi\omega_0t + \frac{j\pi}{5})
\label{eq:u_sin}
\end{equation}
for $j=1,\ldots,5$ where $A=0.5$. 
The rationale for choosing such control input is to explore the 
state-action space, which is essential for convergence of the Q-learning ~\cite{bertsekas1996neuro}.



\begin{figure}[t]
	\centering
	\includegraphics[width=0.9\columnwidth,keepaspectratio=true]
	{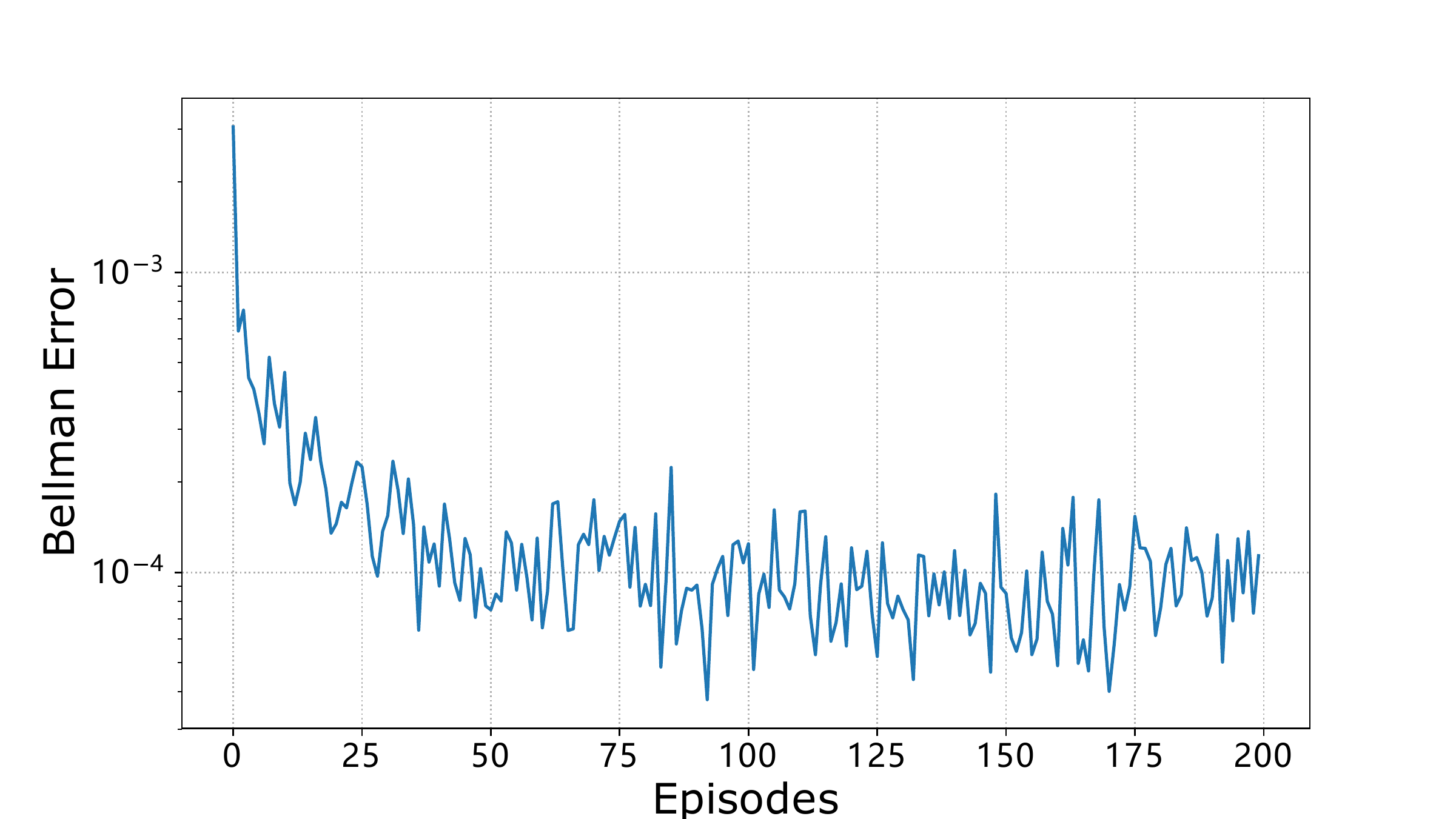}
	\caption{Summary of the Q-learning algorithm result: Average Bellman error defined in~\eqref{eq:BE-mean} versus episode number.}
	\label{fig:BE}
\end{figure}

The $L^2$-norm of the point-wise Bellman error~\eqref{eq:Bellman-error}, averaged  
over the $j$-th episode, is defined according to
\begin{equation}
	e_j := 
	\frac{1}{n_TT}
	\int_{(j-1)n_TT}^{jn_TT}\Big|\mathcal{E}(\thetaN(t),u(t);w(t))\Big|^2 \ud t
	\label{eq:BE-mean}
\end{equation}
The average Bellman error $e_j$ as a function of episode is depicted in Figure~\ref{fig:BE}.  
The decrease in the Bellman error implies that the algorithm is able to
learn the Hamiltonian function that solves the approximate dynamic programming
fixed-point equation~\eqref{eq:g-dynamicsixed-pt}.

\begin{figure*}[t]
	\centering
	\begin{tabular}{cc}
		\subfigure[]{
			\includegraphics[width=0.8\columnwidth,keepaspectratio=true]
			{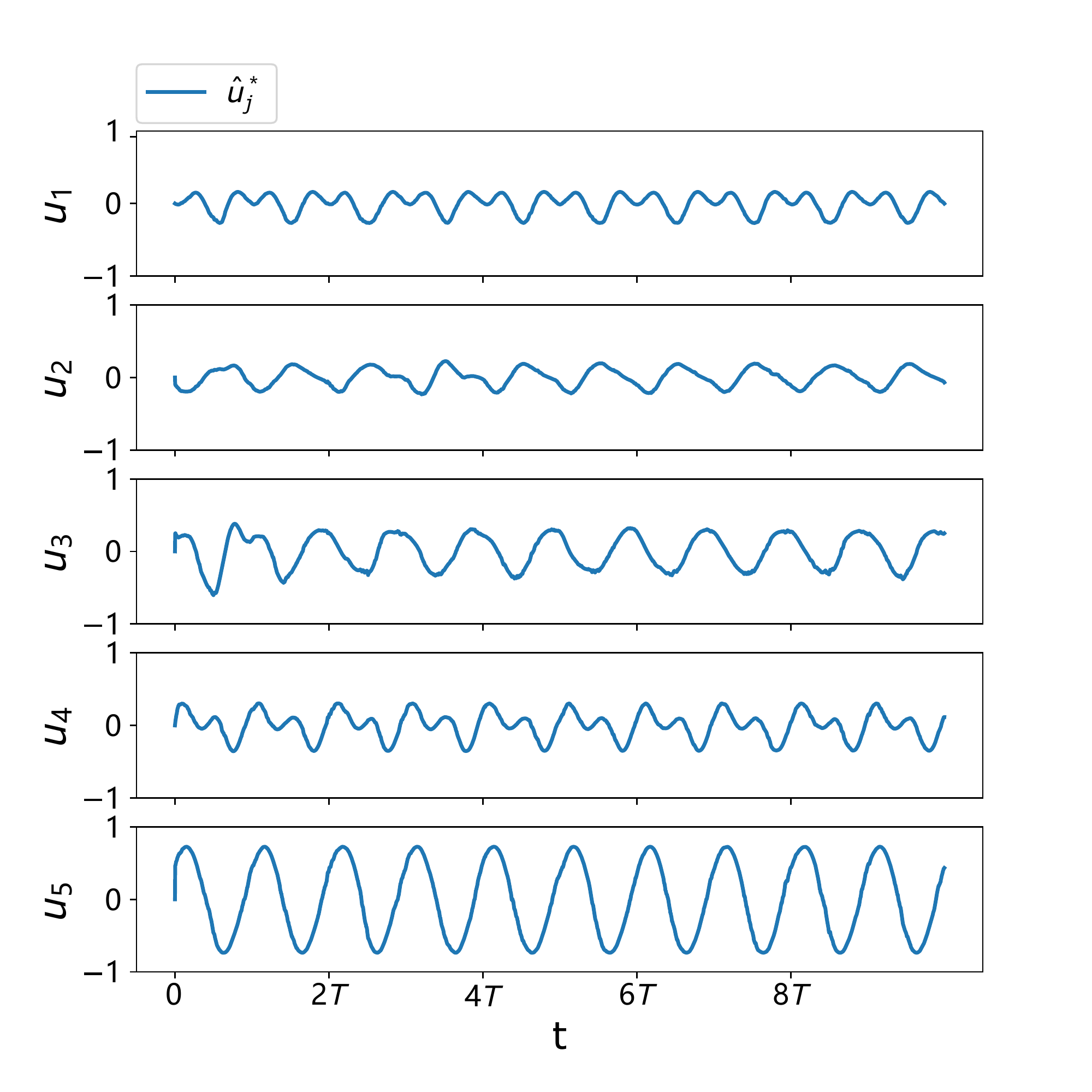}
			\label{fig:u}
		}
		\subfigure[]{
			\includegraphics[width=0.8\columnwidth,keepaspectratio=true]
			{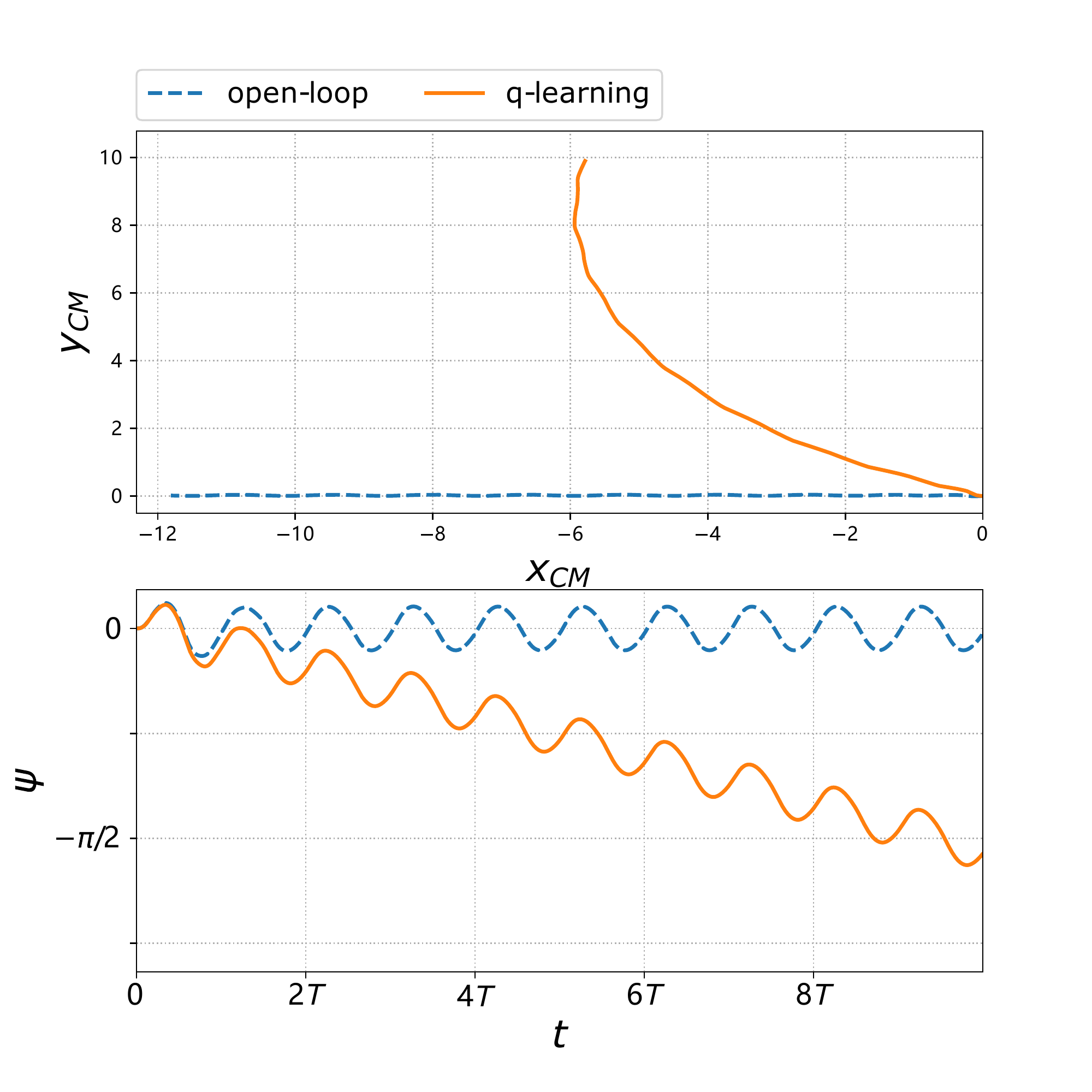}
			\label{fig:g-variables}
		}
	\end{tabular}
	\caption{Summary of control results: 
		(a) The learned control input~\eqref{eq:u-distributed} 
			 from the Q-learning algorithm; 
		(b) Time trace of the global displacement $r\cm=(x\cm,y\cm)$ and the global orientation $\psi$  in open-loop manner
	($u(t)=0$), 
			and using the learned control input.}
	\label{fig:u & g}
\end{figure*} 


Figure~\ref{fig:u} depicts the learned control input $\hat{u}^*(\theta^{(N)}(t),w^*)$ evaluated according to~\eqref{eq:u-distributed} . 
Figure~\ref{fig:g-variables} depicts the resulting global displacement $r\cm(t)$ and the  
global orientation $\psi(t)$, driven with the learned control input. 
It is observed that the learned  control input induces net change in the global 
orientation and turn the snake robot clockwise. 
\section{Conclusions and Future Work}
\label{sec:Conclusions}
%

A bio-inspired framework for learning a sensorimotor 
control of locomotion is introduced and illustrated with a planar coupled rigid body model of a snake robot. 
The framework does not require knowledge of the explicit form of the dynamics and the
observation models.

Although the filtering and control  are implemented in a distributed manner, the Q-learning algorithm is centralized. A possible direction of future work is to implement the learning in a distributed way, so that the overall architecture becomes fully distributed.  Another direction for future work is to  extend the current framework to continuum rod type of models, motivated by applications in soft-robotics.

\appendix

\subsection{Derivation of the dynamic model}
\label{apdx:dynamics}
The dynamic equations are derived from Lagrangian mechanics approach. 
The Lagrangian $L=E(q,\dot{q})-V(q)$ is the difference between kinetic energy $E(q,\dot{q})$ and potential energy $V(q)$, given by:
  \begin{equation*}
  E(q,\dot{q}) = \frac{1}{2}m\dot{r}\cm^2 +  
   \frac{1}{2}\dot{q}^TI_{(q)}\dot{q}~, \quad
   V(q)= \sum_{j=1}^{n-1}
  \frac{1}{2}\kappa_j(q_j-q_{j+1})^2
  \label{eq:E_k,V}
  \end{equation*}
where $m=\sum_{i=j}^{n}m_j$ is the total mass of the system, and $I_{(q)}$ is the inertia matrix.
The Euler-Lagrange equation is, 
\begin{equation}
\frac{\ud}{\ud t}(\frac{\partial L}{\partial \dot{q}}) - \frac{\partial L}{\partial q} = \fgen
\label{eq:Eu-Lag}
\end{equation}
where $\fgen \in \mathbb{R}^{n+2}$ are the generalized forces. 
Generalized forces are defined by $\delta W =  \delta q \fgen$
where $\delta W$ is the virtual work done by nonconservative forces, under infinitesimal variation $\delta q$. Nonconservative forces include actuator torques, viscous friction at each joint, and friction force with surface. 

\begin{table}[tb]
\medskip
\centering
\caption{Model Parameters for the n-link system}
\begin{tabular}{cc}
Parameter & Description \\
\hline
\hline\noalign{\smallskip}
$m_j$ & Mass of link $j$  \\
$J_j$ & Moment of inertia of link $j$  \\
2$l_j$ & Length of link $j$ \\
$c_{t,j}$ &  friction coefficient tangent to link $j$\\
$c_{n,j}$ & friction coefficient normal to link $j$ \\
\hline\noalign{\smallskip}
$\kappa_j$ & Torsional spring coefficient at joint $j$\\
$\zeta_j$ & Viscous friction coefficient at joint $j$\\ 
\hline\noalign{\smallskip}
$\tau_j$ & Input torque amplitude at joint $j$  \\
$\omega_0$ & Input torque frequency  \\
$\sigma_w$ & Noise process std. dev. \\
$\epsilon$ & Control penalty parameter\\
\hline
\\
\multicolumn{2}{c}{Numerical values}\\
\hline
\hline\noalign{\smallskip}
\multicolumn{2}{c}{$
m_j=1.0,\quad J_j=1/3, \quad l_j=1.0,\quad c_{t,j}=0.1,\quad c_{n,j}=0.5$ 
}\\\noalign{\smallskip}
\multicolumn{2}{c}{$
\kappa_j=3.0,\quad \zeta_j=0.1$ for $j=1,2,3,4$ 
}\\\noalign{\smallskip}
\multicolumn{2}{c}{$
\tau_0=[2.0,1.1,1.0,2.0],\quad \omega_0=1.0, \quad \sigma_w^2=0.1,\quad \epsilon=10.0$ 
}\\\noalign{\smallskip}
\hline
\end{tabular}
\label{tab:n-linkParam-1}
\end{table}

Considering generalized forces, the equations of motion are succinctly expressed as, 
 \begin{equation}
 \begin{aligned}
 I_{(q)}\ddot{q} + \C \dot{q}^2 + \tilde{\kappa} q &= D^T \tau  -\tilde{\zeta} \dthet - 
 \CC_{q q}\dthet -
 \CC_{q r}\dot{r}\cm \\
 \frac{\ud}{\ud t}
(m\dot{r}\cm)&=  -\CC_{rr}\dot{r}\cm - \CC_{qr}^T\dot{q}
 \end{aligned}
 \label{eq:q-dynamics}
 \end{equation}
 where $\tau=\lbrack \tau_1,\ldots,\tau_{n-1}\rbrack$ are the actuator torques, $\tilde{\kappa}$ is the stiffness matrix, and $\tilde{\zeta}$ is the friction coefficient matrix. 
 The terms involving $\CC_{qq},\CC_{qr},\CC_{rr}$ arise due to friction with the surface.
 The matrix $D \in \mathbb{R}^{n-1\times n}$ is the difference operator.
 These parameters are tabulated in Table~\ref{tab:n-linkParam-2}. A detailed derivation of the equations of motion appears in \cite{Iwasaki-snake}.

\begin{table}[tb]
\medskip
\centering
\caption{Auxiliary Parameters for the n-link system}
\begin{tabular}{cc}
\hline
\hline
\noalign{\smallskip}
\multicolumn{2}{c}{
$M=\text{diag}(m_j),
\quad J=\text{diag}(J_j),
\quad L=\text{diag}(l_j),
$}\\\noalign{\smallskip}
\multicolumn{2}{c}{
$
C\NN=\text{diag}(c_{n,j}),\quad
C\TT=\text{diag}(c_{t,j})
$}\\\noalign{\smallskip}
\multicolumn{2}{c}{
$ \lbrack D\rbrack_{n-1\times n} ~~s.t~~ [Dx]_j=x_j-x_{j+1}, \quad 
 \lbrack A\rbrack_{n-1\times n} ~~s.t~~ [Ax]_j=x_j+x_{j+1}
$}\\\noalign{\smallskip}
\multicolumn{2}{c}{
$D^{+}=D^T(DD^T)^{-1}\quad e=[1,\ldots,1]^T
$}\\\noalign{\smallskip}
\multicolumn{2}{c}{
$\kappa=\text{diag}(\kappa_j),\quad
\zeta=\text{diag}(\zeta_j),\quad
\tilde{\kappa}=D^T\kappa D,\quad
\tilde{\zeta}=D^T\zeta D
$}\\\noalign{\smallskip}
\multicolumn{2}{c}{
$H = LA^T(DM^{-1}D^T)^{-1}AL,\quad
B = M^{-1}D^T(DM^{-1}D^T)^{-1}AL
$
}\\\noalign{\smallskip}
\hline\noalign{\smallskip}
\multicolumn{2}{c}{
 $c_q=\cos(q_j),
\quad s_q=\sin(q_j)  
$}\\\noalign{\smallskip}
\multicolumn{2}{c}{
$\lbrack{I_{(q)}}\rbrack_{ij}=H_{ij}\cos(q_i-q_j) + J_{ij}\quad 
\lbrack{\C}\rbrack_{ij}=H_{ij}\sin(q_i-q_j)
$}\\\noalign{\smallskip}
\hline\noalign{\smallskip}
\multicolumn{2}{c}{$
\lbrack{B_s}\rbrack_{ij}=B_{ij}\sin(q_i-q_j)\quad 
\lbrack{B_c}\rbrack_{ij}=B_{ij}\cos(q_i-q_j)
$}\\\noalign{\smallskip}
\multicolumn{2}{c}{$
\CC_{q q} = B_s^TMC\TT B_s + B_c^TMC\NN B_c + C\NN J 
$}\\\noalign{\smallskip}
\multicolumn{2}{c}{$
\CC_{q v_1} = B_s^TMC\TT c_{q-\psi} -  B_c^TMC\NN s_{q-\psi}
$}\\\noalign{\smallskip}
\multicolumn{2}{c}{$
\CC_{q v_2} = B_s^TMC\TT s_{q-\psi} +  B_c^TMC\NN c_{q-\psi}
$}\\\noalign{\smallskip}
\multicolumn{2}{c}{$
\CC_{vv} = c_{q-\psi} ^TMC\TT c_{q-\psi} + s_{q-\psi} ^TMC\NN s_{q-\psi} 
$}\\\noalign{\smallskip}
\multicolumn{2}{c}{$
\CC_{v_2v_2} = s_{q-\psi} ^TMC\TT s_{q-\psi} + c_{q-\psi} ^TMC\NN c_{q-\psi} 
$}\\\noalign{\smallskip}
\multicolumn{2}{c}{$
\CC_{v_1v_2} = s_{q-\psi} ^TM(C\TT-C\NN) c_{q-\psi} 
$}\\\noalign{\smallskip}
\multicolumn{2}{c}{$
\CC_{qr} = \CC_{qv}R^T(\psi),\quad
\CC_{rr} = R(\psi)\CC_{vv}R^T(\psi) 
$}\\\noalign{\smallskip}
\hline
\end{tabular}
\label{tab:n-linkParam-2}
\end{table}

\noindent
{\bf Shape dynamics:} 
The coordinate transformation  $(x,\psi) \leftrightarrow q$ is given by 
\begin{equation}
\begin{bmatrix}
x\\
\psi
\end{bmatrix}
=
\begin{bmatrix}
D\\
\frac{1}{n}e^T
\end{bmatrix}
q
\Rightarrow
q
=
\begin{bmatrix}
D^+ & e
\end{bmatrix}
\begin{bmatrix}
x\\
\psi
\end{bmatrix}
\label{eq:q-g}
\end{equation}
where $e=[1,\ldots,1]^T \in \mathbb{R}^n$, and $D^+=D^T(DD^T)^{-1}$. 
Then the dynamic equation for the shape variable $x$ is,
\begin{equation}
\begin{aligned}
\ddot{x}&=(DI_{(x) }^{-1}D^T)(\tau - \kappa x-\zeta \dot{x})\\
&+DI_{(x) }^{-1}(-\C\dot{q}^2-\CC_{qq}D^+\dot{x} - \CC_{qq}e\dot{\psi} - \CC_{qv}v)
\end{aligned}
\label{eq:i-dynamics}
\end{equation}
where $I_{(x)}:=I_{(q(x))}$. 
This is the explicit form of the second order ode in~\eqref{eq:i-dynamics-u}.

\noindent
{\bf Group dynamics:~} Define the  total angular momentum $\mu=e^TI_{(q)}\dot{q}$, and the group velocity $v=R(\psi)^T\dot{r}\cm$
where $R(\psi)=
\begin{bmatrix}
\cos(\psi) & -\sin(\psi) \\
\sin(\psi) & \cos(\psi)
\end{bmatrix}$ is the rotation matrix. 
The dynamics for these two variables are
\begin{equation}
\begin{aligned}
\frac{\ud \mu}{\ud t}&=-e^T\CC_{qq}e\dot{\psi}-e^T\CC_{qv}v-e^T\CC_{qq}{D}^{+}\dot{x} \\
\frac{\ud mv}{\ud t} &=- m
\begin{bmatrix}
0 & -\dot{\psi}\\
\dot{\psi} & 0
\end{bmatrix}
 v -\CC_{vv}v - \CC_{vq}e\dot{\psi}-\CC_{vq}{D}^{+}\dot{x}
\end{aligned}
\label{eq:momentum}
\end{equation}
Assuming the inertial terms are negligible, a first order ode is obtain for the evolution of the group variables $(r\cm,\psi)$:
 \begin{equation}
 \frac{\ud}{\ud t}
 \begin{bmatrix}
	\psi \\
	R(\psi)^T r\cm
 \end{bmatrix}=-
 \begin{bmatrix}
 e^T\CC_{q q}e & e^T\CC_{qv} \\
 \CC_{vq}e & \CC_{vv} \\
 \end{bmatrix}^{-1}
 \begin{bmatrix}
 e^T \CC_{q q} \\
 \CC_{vq}
 \end{bmatrix}
 {D}^{+}\dot{x} 
  \label{eq:g-dynamics}
 \end{equation}
 This is the first order ode that appears in~\eqref{eq:g-dynamics-u}.

\bibliographystyle{plain}
\bibliography{reference}

\end{document}